\documentclass[journal]{IEEEtran}
\usepackage[caption=false,font=footnotesize]{subfig}
\usepackage[dvips]{graphicx}
\usepackage{epsfig}
\usepackage{subfig}
\usepackage{amsmath}
\usepackage{amssymb}

\usepackage{multirow}

\usepackage{color}
\usepackage[normalem]{ulem}

\begin{document}

\title{A steep-slope MoS$_2$-nanoribbon MOSFET based on an intrinsic cold-contact effect}

\author{Demetrio~Logoteta,  
     	Marco~G.~Pala,~\IEEEmembership{Member,~IEEE},
        Jean~Choukroun,
        Philippe~Dollfus,~\IEEEmembership{Member,~IEEE}, and
        Giuseppe~Iannaccone,~\IEEEmembership{Fellow,~IEEE}
\thanks{This work has been partially supported under the European Unions Horizon 2020 research and innovation programme (grant agreement QUEFORMAL no. 829035) and under MIUR PRIN program 
(project FIVE2D).
}
\thanks{D.~Logoteta is with Dipartimento di Ingegneria dell'Informazione, Universit\'a di Pisa, Via G. Caruso 16, 56126 Pisa, Italy, and with Centre de Nanosciences et de Nanotechnologies, Centre National de la Recherche Scientifique, Universit\'e Paris-Sud, Universit\'e Paris-Saclay, 91120 Palaiseau, France (e-mail: demetrio.logoteta@u-psud.fr)}
\thanks{J. Choukroun, M.~G.~Pala and P.~Dollfus are with Centre de Nanosciences et de Nanotechnologies, Centre National de la Recherche Scientifique, Universit\'e Paris-Sud, Universit\'e Paris-Saclay, 91120 Palaiseau, France.}
\thanks{G.~Iannaccone is with Dipartimento di Ingegneria dell'Informazione, Universit\'a di Pisa, Via G. Caruso 16, 56126 Pisa, Italy}
}

\maketitle

\begin{abstract}
We propose a steep-slope MoS$_2$-nanoribbon field-effect transistor that exploits a narrow-energy conduction band to intrinsically filter out the thermionic tail of the electron energy distribution. We study the device operation principle and the performance dependence on the design parameters through atomistic self-consistent quantum simulations. Our results indicate that the device can provide high ${\bf I_{\rm ON}/I_{\rm OFF}}$ ratios, compatible with electronic applications, albeit biased at ultra-low voltages of around 0.1~V. 
\end{abstract}


\IEEEpeerreviewmaketitle

\section{Introduction}
\label{Sec:int}

\IEEEPARstart{S}{teep}-slope transistors, namely transistors able to break the 60 mV/dec thermal limit of the subthreshold swing (SS), have attracted substantial interest as energy-efficient beyond-CMOS switches. 
Several physical mechanisms, including energy filtering {\it via} tunneling~\cite{Seabaugh_IEEEproceedings_2010}, impact ionization~\cite{gopalakrishnan2005} and negative gate capacitance~\cite{salahuddin2008use}, have been proposed as possible options to design transistors with subthermionic SS. 

Recently, the rise of 2D materials has introduced further opportunities, relying on the possibility to exploit taylored band structures~\cite{qiu2018dirac}, obtained both by choosing in a wide portfolio of unary or compound materials~\cite{mounet2018two}, and by effectively tuning their properties by lateral confinement, modification of the geometry and the passivation of edges, combination in heterostructures~\cite{iannaccone2018quantum,Cao2016operation}, strain or electrical gating. 
      
In this paper, we propose a steep-slope transistor based on an armchair nanoribbon of MoS$_2$ with dangling bonds passivated by hydrogen atoms. As in traditional tunnel field-effect transistors (TFETs), the subthermionic behavior is obtained by cutting-off the Boltzmann tail of the current. However, while in TFETs this is achieved by inducing a band-to-band tunneling, which usually severely degrades the ON current, the proposed device exploits a lack of available states, intrinsically resulting from the nanoribbon band structure. This translates to a filtering effect analogous to that which could be induced by cooling down both the source and drain regions~\cite{liu2018first}. Compared to other recently proposed cold-source architectures~\cite{liu2018first}, the device does not require heterojunctions, and thus avoids the detrimental effect of defects at the interface between different materials.


\section{Device and Model}
\label{Sec:mod}

\begin{figure}[!t]
    	\centering
		\includegraphics[width=0.77\columnwidth]{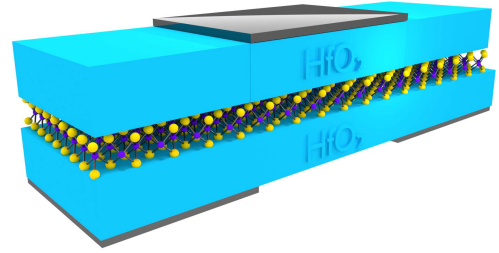}
		\caption{\protect \footnotesize                 
        Sketch of the simulated MoS$_2$ nanoribbon-based transistor.
        }
		\label{Fig:Sketch}  
\end{figure}

\begin{figure}[!t]
    	\centering
		\includegraphics[width=0.97\columnwidth]{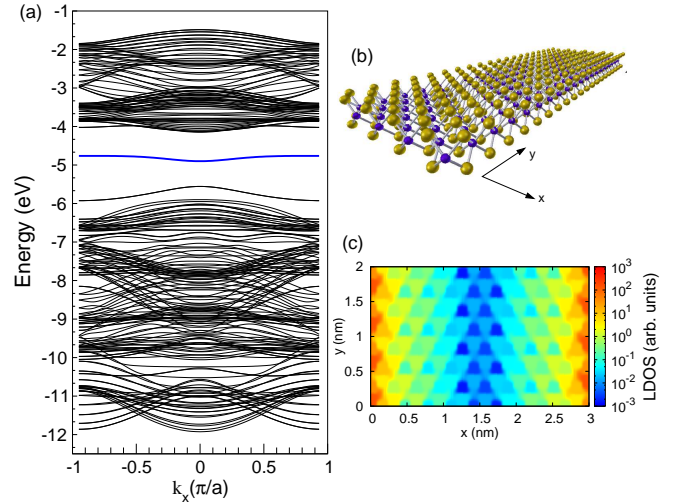}
		\caption{\protect \footnotesize                 
        (a) Band structure of the 20-dimer wide, hydrogen-passivated MoS$_2$ armchair nanoribbon. The 
        isolated band is highlighted in bold blue. 
        (b) Atomic structure of the nanoribbon. (c) Color map of the local density of states at the 
        minimum of the bold blue bands in panel (a).
        }
		\label{Fig:BS}  
\end{figure}

The device geometry is sketched in Fig.~\ref{Fig:Sketch}. A 20-dimer wide ($\sim$3.2~nm) MoS$_2$ armchair nanoribbon with hydrogen-passivated edges is stacked between 3~nm thick HfO$_2$ layers. Close to the source and drain contacts the nanoribbon is electrostatically doped by means of two back gates, 5 and 9~nm long, respectively. The source contact extension has been shortened with respect to the drain counterpart for numerical reasons, in order to help the potential stabilization and facilitate convergence. Throughout the paper, the back gates are held at the same voltage of 1.2~V, corresponding to an n-type doping of $\sim 1.6\times 10^6$~cm$^{-1}$. The transistor is controlled with a 15~nm long top gate, unless otherwise stated.   

The MoS$_2$ nanoribbon is described by the 22-orbital tight-binding Hamiltonian formulated in Ref.~\cite{choukroun2018high}, obtained by projection of the corresponding density functional Hamiltonian on a basis of maximally localized Wannier functions~\cite{fang2015ab}. The ballistic transport through the device is simulated within the non-equilibrium Green's function formalism, by self-consistently coupling the transport and Poisson equations on a finite element grid. 

The band and lattice structure of the nanoribbon is reported in Fig.~\ref{Fig:BS}, in panels (a) and (b), respectively. As previous works have pointed out~\cite{pan2012edge,yue2012bandgap}, the lateral confinement results in the appearance within the band gap of MoS$_2$ of states localized at the nanoribbon edges. Here, we will focus in particular on the couple of almost degenerate, isolated narrow bands plotted in bold blue in Fig.~\ref{Fig:BS} (a). 
The energy width of these bands is smaller than 0.14 eV and they are more than 0.6 eV apart both from above and underlying bands. 
Fig.~\ref{Fig:BS} (c) provides a map of the local density of states (LDOS) on the nanoribbon plane at the energy corresponding to the minimum of these bands and demonstrates the exponential localization at the edges. As far as states on the opposite edges are well separated, these bands and their distance from the neighboring ones are independent of the nanoribbon width. Sizable modifications, particularly an increase of the band splitting, only occurs for sub-7 dimers widths.  

According to {\it ab-initio} computations~\cite{pan2012edge}, the nanoribbon is semiconducting, with the Fermi level lying in the gap between the isolated bands and the hole-like bands immediately below {(see Fig.~\ref{Fig:BS}(a))}. 
If a current flow through the selected bands is enabled, the energy spectrum of the current will be nonzero only in a narrow window equal to the band width. In a MOSFET architecture, this entails a suppression of the spectral components of the current through the source and drain contacts at energies higher than the upper edge of the bands. The flatness of the bands and their isolation from other bands, therefore, intrinsically induce a cold-contact effect.   



\section{Results and discussion}
\label{Sec:res}

\begin{figure}[t]
    	\centering
		\includegraphics[width=0.97\columnwidth]{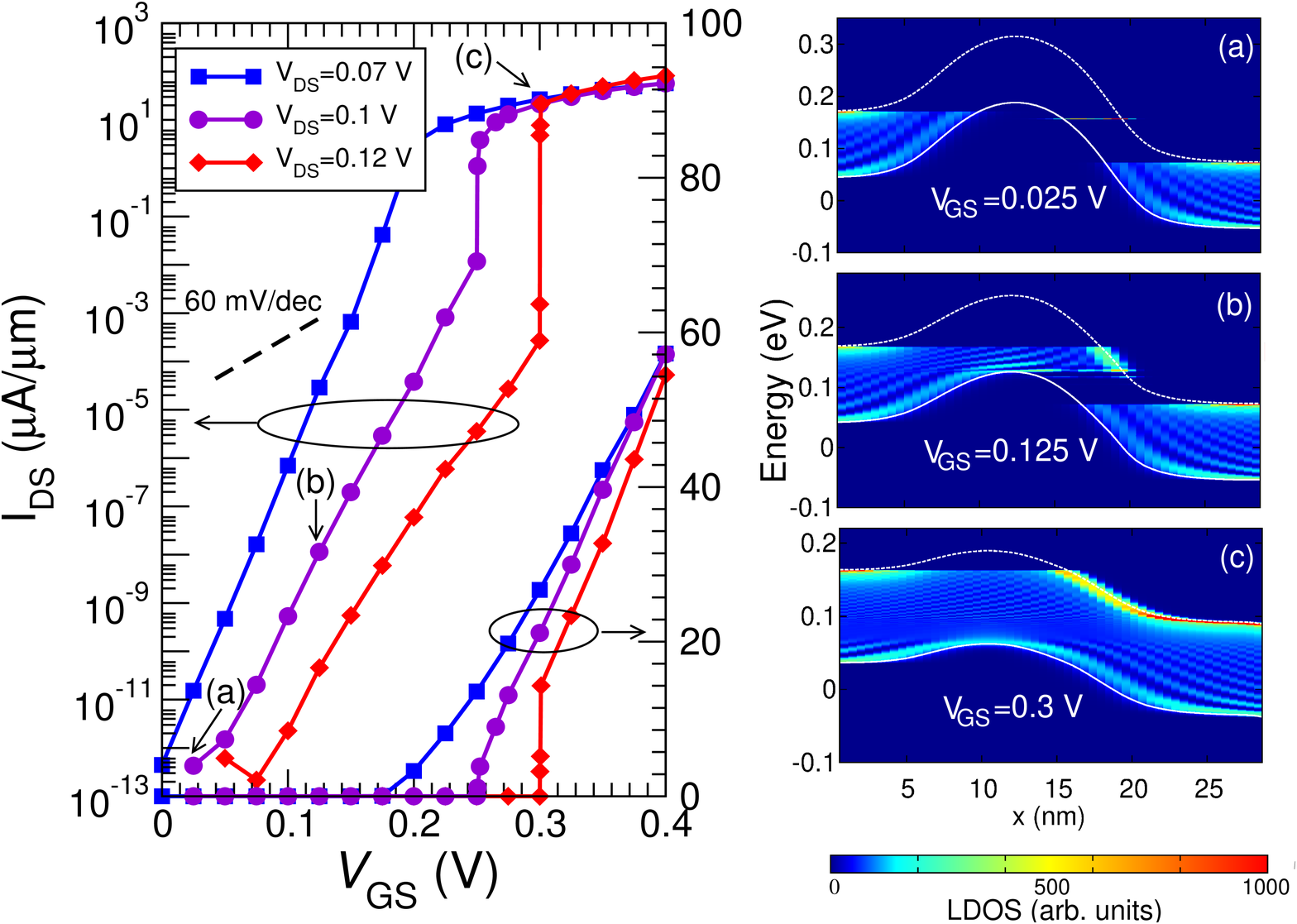}
		\caption{\protect \footnotesize                 
        Left: transfer characteristics of the device for $V_{\rm DS}=$0.07, 0.1 and 0.12~V. Right: local 
        density of states along the transport direction $x$ for $V_{\rm DS}=$0.1~V. The solid
        and dashed white lines denote the bottom and upper edge of the conducting band, respectively. 
        Panels (a), (b) and (c) refer to the $V_{\rm GS}$ value indicated by the same letter in the left  
        panel.
        }
		\label{Fig:IV_VDS}  
\end{figure}

The left panel of Fig.~\ref{Fig:IV_VDS} illustrates the transfer characteristics of the device for three different values of the source-drain bias $V_{\rm DS}$. As the transport occurs {\it via} edge states, the current does not scale with the nanoribbon width. Accordingly, here and in the following the current is normalized by assuming that an array of 150 parallel-connected devices is arranged over a width of 1 $\mu$m. Dense arrays of graphene nanoribbons with a similar pitch have been reported in Ref.~\cite{Abbas2014patterning}. 

We first focus on the curve for $V_{\rm DS}=0.1$~V. To gain physical insights into the operation of the device, we report in the right panel of Fig.~\ref{Fig:IV_VDS} the local density of state along the transport direction for the values of the top-gate voltage $V_{\rm GS}$ indicated with (a), (b) and (c) in the left panel. 
At small enough $V_{\rm GS}$ (case (a)), the top of the channel barrier is higher than the highest energy at which electron can be injected from the source contact. Therefore, all the electrons impinging on the barrier are back reflected. As the barrier lowers for increasing $V_{\rm GS}$ (case (b)), electrons start to be transmitted over the barrier. However, they still cannot reach the drain contact beacuse no states at the same energy are available on the drain side. Electrons are therefore entirely back-scattered toward the source when they reach  the upper edge of the band (dashed curves in the figure). In the cases (a) and (b), the leakage current through the device is only due to tunneling. This current is effectively modulated by the top gate, which results in a subthermionic SS of $\sim 22$~mV/dec. At large enough $V_{\rm GS}$ (case (c)), the top of the barrier is pushed below the upper edge of the band at the drain contact, therefore establishing an overlap window between available states at source and drain and enabling thermionic conduction. The threshold value of $V_{\rm GS}$ beyond which the device turns on is around 0.25~V and is associated to the sub-1~mV/dec subthreshold swing segment of the transfer characteristic. The apparent decrease of $V_{\rm DS}$ for increasing $V_{\rm GS}$ in the maps is due to the rearrangement of the self-consistent potential required to maintain the charge neutrality at the contacts~\cite{Bescond20043D}.

It can be easily argued from the above discussion that an energy overlap between states on the source and the drain side can be induced only for $V_{\rm DS}$ smaller than the energy width of the conducting band. For larger $V_{\rm DS}$ the device cannot be turned on in ballistic conditions. We also underline that the subthermionic operation of the device in ballistic conditions is driven by the effective cooling-down of the drain, while the energy filtering at source does not play any essential role. 

\begin{figure}[!t]
    	\centering
		\includegraphics[width=0.9\columnwidth]{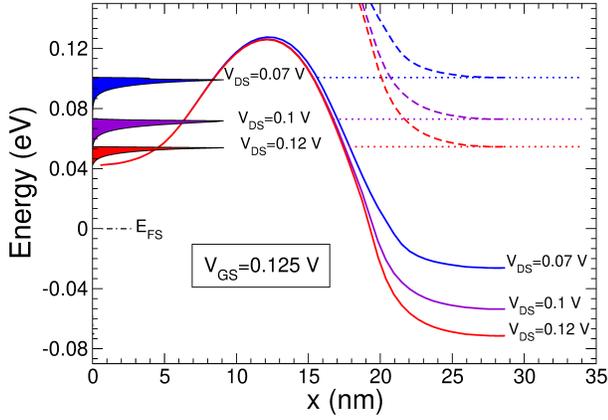}
		\caption{\protect \footnotesize                 
        Bottom (solid lines) and top (dashed lines) of the conduction band for $V_{\rm DS}=$0.07, 0.1 and 
        0.12 V, and $V_{\rm GS}=$0.125~V. The corresponding current spectra, normalized to the same value  
        at their maximum, are also shown. Dotted lines are guides for the eye. The source Fermi level is 
        $E_{\rm FS}=0$. 
        }
		\label{Fig:Subbands}  
\end{figure}

The evolution of the transfer characteristics as a function of $V_{\rm DS}$ can be understood from Fig.~\ref{Fig:Subbands}, illustrating the bottom of the conduction band and the corresponding current spectrum for the three considered values of $V_{\rm DS}$ and for $V_{\rm GS}=$0.125~V. The plot shows that the tunneling current spectrum always peaks at the upper edge of the band at the drain contact (dotted lines). As $V_{\rm DS}$ increases, the tunneling window moves at lower energies and electrons face a thicker barrier. As a consequence, the tunneling component of the current decreases and a larger jump in the transfer characteristics occurs when the regime switches from tunneling-dominated to predominantly thermionic. The SS degradation for increasing $V_{\rm DS}$ when the transport is tunneling-dominated originates from the decreasing sensitivity of the tunneling current to the barrier modulation, as electrons are injected closer to the base of the barrier. Finally, the increase of the threshold voltage with $V_{\rm DS}$ is a direct consequence of the corresponding downshift of the upper edge of the band at the drain contact with respect to the top of the barrier.


\begin{figure}[!t]
    	\centering
		\includegraphics[width=0.95\columnwidth]{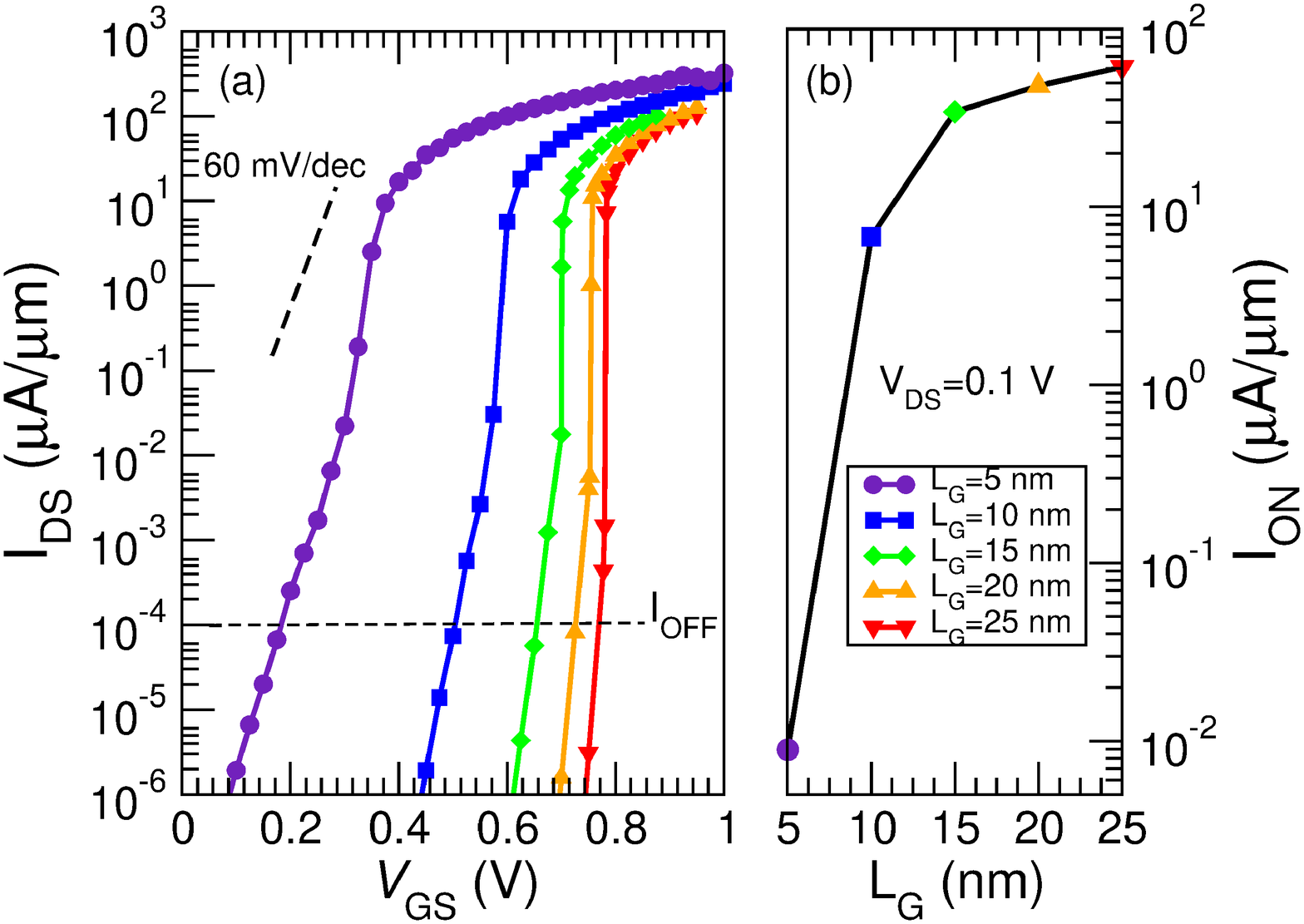}
		\caption{\protect \footnotesize                 
        (a) Transfer characteristics of the device for $V_{\rm DS}=$0.1~V and for several gate lengths, 
        ranging from 5 to 25~nm; (b) On-current as a function of the gate length. The OFF-current value 
        is set at $10^{-4} \mu$A/$\mu$m~\cite{IRDS} (dashed line in panel (a)). $I_{\rm ON}$ is computed  		as $I_{\rm ON} = I_{\rm DS}(V_{\rm GS}^{\rm off}+V_{\rm DS})$~\cite{IRDS}, where $V_{\rm GS}^{\rm 		off}$ is defined by $I_{\rm DS}(V_{\rm GS}^{\rm off})=I_{\rm OFF}$.
        }
		\label{Fig:IV_LG}  
\end{figure}

The scaling behavior of the device is illustrated in Fig.~\ref{Fig:IV_LG}, where the device transfer characteristics (panel (a)) and the corresponding on-current values (panel (b)) are reported for several top-gate lengths. Although the SS remains subthermonic also for the ultra-scaled gate length of 5 nm, it degrades as $L_{\rm G}$ decreases, due to the increase of the tunneling component of the current. This degradation is particularly significant for $L_{\rm G}$ scaling from 15~nm to 5~nm, which corresponds to a drop of the $I_{\rm ON}/I_{\rm OFF}$ ratio of more than three orders of magnitude.  
The evolution of transfer characteristics with $L_{\rm G}$ around the turning-on threshold is similar to their dependence on $V_{\rm DS}$, according to the fact that changes are driven in both cases by modulations of the leakage tunneling current.   

We conclude by discussing some aspects not directly taken into account in our simulations.
Electron-phonon scattering is not included in our model, due to the lack in the literature of phonon and electron-phonon interaction parameters for the specific case at hand. Phonon absorption and emission can assist electrons in overcoming the channel barrier and relax toward available states at the drain, respectively, in cases where it would not be allowed in the ballistic regime (for instance, in the conditions (b) and (c) in the right panel of Fig.~\ref{Fig:IV_VDS}). This is likely to degrade the subtrheshold swing of the device~\cite{Pala2016}, especially at the onset of conduction. Neverthless, the device is expected to maintain subthermionic capabilities, due to the energy filtering effect at the source contact that, differently from the drain case, is not influenced by electron-phonon interactions. 

The device requires small fabrication tolerances, as large edge roughness can suppress the transmission~\cite{park2018impact}. While advances in fabrication techniques indicate that defectless MoS$_2$ nanoribbons down to sub-nanometer width can be manufactured~\cite{liu2013top}, we also point out that hydrogen-passivated MoS$_2$ armchair nanoribbons can be more generally used as cold-source injectors in composite structures based on planar~\cite{iannaccone2018quantum} or van der Waals heterojunctions~\cite{Cao2016operation}. In these devices, the nanoribbon can be in principle shortened below the edge roughness correlation length. Particularly, the variety of possible van der Waals heterojunctions allowed by the absence of lattice matching constraints leaves open many opportunities of optimization and material-device co-design. 


\begin{thebibliography}{10}
\providecommand{\url}[1]{#1}
\csname url@samestyle\endcsname
\providecommand{\newblock}{\relax}
\providecommand{\bibinfo}[2]{#2}
\providecommand{\BIBentrySTDinterwordspacing}{\spaceskip=0pt\relax}
\providecommand{\BIBentryALTinterwordstretchfactor}{4}
\providecommand{\BIBentryALTinterwordspacing}{\spaceskip=\fontdimen2\font plus
\BIBentryALTinterwordstretchfactor\fontdimen3\font minus
  \fontdimen4\font\relax}
\providecommand{\BIBforeignlanguage}[2]{{%
\expandafter\ifx\csname l@#1\endcsname\relax
\typeout{** WARNING: IEEEtran.bst: No hyphenation pattern has been}%
\typeout{** loaded for the language `#1'. Using the pattern for}%
\typeout{** the default language instead.}%
\else
\language=\csname l@#1\endcsname
\fi
#2}}
\providecommand{\BIBdecl}{\relax}
\BIBdecl

\bibitem{Seabaugh_IEEEproceedings_2010}
A.~Seabaugh and Q.~Zhang, ``{Low-Voltage Tunnel Transistors for Beyond CMOS
  Logic},'' \emph{Proceedings of the IEEE}, vol.~98, no.~12, pp. 2095 --2110,
  Dec 2010, doi: \url{10.1109/JPROC.2010.2070470}.

\bibitem{gopalakrishnan2005}
K.~Gopalakrishnan, P.~B. Griffin, and J.~D. Plummer, ``{Impact ionization MOS
  (I-MOS)-Part I: device and circuit simulations},'' \emph{IEEE Transactions on
  electron devices}, vol.~52, no.~1, pp. 69--76, Dec 2005, doi:
  \url{10.1109/TED.2004.841344}.

\bibitem{salahuddin2008use}
S.~Salahuddin and S.~Datta, ``Use of negative capacitance to provide voltage
  amplification for low power nanoscale devices,'' \emph{Nano letters}, vol.~8,
  no.~2, pp. 405--410, Dec 2008, doi: \url{10.1021/nl071804g}.

\bibitem{qiu2018dirac}
C.~Qiu, F.~Liu, L.~Xu, B.~Deng, M.~Xiao, J.~Si, L.~Lin, Z.~Zhang, J.~Wang,
  H.~Guo \emph{et~al.}, ``Dirac-source field-effect transistors as
  energy-efficient, high-performance electronic switches,'' \emph{Science},
  vol. 361, no. 6400, pp. 387--392, Jul 2018, doi:
  \url{10.1126/science.aap9195}.

\bibitem{mounet2018two}
N.~Mounet, M.~Gibertini, P.~Schwaller, D.~Campi, A.~Merkys, A.~Marrazzo,
  T.~Sohier, I.~E. Castelli, A.~Cepellotti, G.~Pizzi \emph{et~al.},
  ``Two-dimensional materials from high-throughput computational exfoliation of
  experimentally known compounds,'' \emph{Nature nanotechnology}, vol.~13,
  no.~3, p. 246, Feb 2018, doi: \url{10.1038/s41565-017-0035-5}.

\bibitem{iannaccone2018quantum}
G.~Iannaccone, F.~Bonaccorso, L.~Colombo, and G.~Fiori, ``Quantum engineering
  of transistors based on 2d materials heterostructures,'' \emph{Nature
  nanotechnology}, vol.~13, pp. 183--191, Mar 2018, doi:
  \url{10.1038/s41565-018-0082-6}.

\bibitem{Cao2016operation}
J.~{Cao}, D.~{Logoteta}, S.~{{\"O}zkaya}, B.~{Biel}, A.~{Cresti}, M.~G. {Pala},
  and D.~{Esseni}, ``Operation and design of van der waals tunnel transistors:
  A 3-d quantum transport study,'' \emph{IEEE Transactions on Electron
  Devices}, vol.~63, pp. 4388--4394, Nov 2016, doi:
  \url{10.1109/TED.2016.2605144}.

\bibitem{liu2018first}
F.~{Liu}, C.~{Qiu}, Z.~{Zhang}, L.~{Peng}, J.~{Wang}, Z.~{Wu}, and H.~{Guo},
  ``{First Principles Simulation of Energy efficient Switching by Source
  Density of States Engineering},'' in \emph{2018 IEEE International Electron
  Devices Meeting (IEDM)}, Dec 2018, pp. 33.2.1--33.2.4, doi:
  \url{10.1109/IEDM.2018.8614597}.

\bibitem{choukroun2018high}
J.~Choukroun, M.~Pala, S.~Fang, E.~Kaxiras, and P.~Dollfus, ``{High performance
  Tunnel Field Effect Transistors based on in-plane transition metal
  dichalcogenide heterojunctions},'' \emph{Nanotechnology}, vol.~30, no.~2, p.
  025201, Nov 2018, doi: \url{10.1088/1361-6528/aae7df}.

\bibitem{fang2015ab}
S.~Fang, R.~K. Defo, S.~N. Shirodkar, S.~Lieu, G.~A. Tritsaris, and E.~Kaxiras,
  ``{{\it Ab initio} tight-binding Hamiltonian for transition metal
  dichalcogenides},'' \emph{Physical Review B}, vol.~92, no.~20, p. 205108, Nov
  2015, doi: \url{10.1103/PhysRevB.92.205108}.

\bibitem{pan2012edge}
H.~Pan and Y.-W. Zhang, ``{Edge-dependent structural, electronic and magnetic
  properties of MoS$_2$ nanoribbons},'' \emph{Journal of Materials Chemistry},
  vol.~22, no.~15, pp. 7280--7290, Mar 2012, doi: \url{10.1039/C2JM15906F}.

\bibitem{yue2012bandgap}
Q.~Yue, S.~Chang, J.~Kang, X.~Zhang, Z.~Shao, S.~Qin, and J.~Li, ``{Bandgap
  tuning in armchair MoS$_2$ nanoribbon},'' \emph{Journal of Physics: Condensed
  Matter}, vol.~24, no.~33, p. 335501, Jul 2012, doi:
  \url{10.1088/0953-8984/24/33/335501}.

\bibitem{Abbas2014patterning}
A.~N. Abbas, G.~Liu, B.~Liu, L.~Zhang, H.~Liu, D.~Ohlberg, W.~Wu, and C.~Zhou,
  ``Patterning, characterization, and chemical sensing applications of graphene
  nanoribbon arrays down to 5 nm using helium ion beam lithography,'' \emph{ACS
  Nano}, vol.~8, pp. 1538--1546, Jan 2014, doi: \url{10.1021/nn405759v}.

\bibitem{Bescond20043D}
M.~{Bescond}, K.~{Nehari}, J.~L. {Autran}, N.~{Cavassilas}, D.~{Munteanu}, and
  M.~{Lannoo}, ``3d quantum modeling and simulation of multiple-gate nanowire
  mosfets,'' in \emph{IEDM Technical Digest. IEEE International Electron
  Devices Meeting, 2004.}, Dec 2004, pp. 617--620, doi:
  \url{10.1109/IEDM.2004.1419237}.

\bibitem{IRDS}
``{International Roadmap for Devices and Systems, IRDS (2017), More Moore
  Tables},'' \url{https://irds.ieee.org/roadmap-2017}, accessed: 2019-03-21.

\bibitem{Pala2016}
M.~G. Pala, C.~Grillet, J.~Cao, D.~Logoteta, A.~Cresti, and D.~Esseni,
  ``{Impact of inelastic phonon scattering in the OFF state of
  Tunnel-field-effect transistors},'' \emph{Journal of Computational
  Electronics}, vol.~15, pp. 1240--1247, Dec 2016, doi:
  \url{10.1007/s10825-016-0900-8}.

\bibitem{park2018impact}
J.~Park, M.~Mouis, F.~Triozon, and A.~Cresti, ``{Impact of edge roughness on
  the electron transport properties of MoS$_2$ ribbons},'' \emph{Journal of
  Applied Physics}, vol. 124, no.~22, p. 224302, Dec 2018, doi:
  \url{10.1063/1.5050383}.

\bibitem{liu2013top}
X.~Liu, T.~Xu, X.~Wu, Z.~Zhang, J.~Yu, H.~Qiu, J.-H. Hong, C.-H. Jin, J.-X. Li,
  X.-R. Wang \emph{et~al.}, ``Top--down fabrication of sub-nanometre
  semiconducting nanoribbons derived from molybdenum disulfide sheets,''
  \emph{Nature communications}, vol.~4, p. 1776, Apr 2013, doi:
  \url{10.1038/ncomms2803}.

\end{thebibliography}

\end{document}